\documentclass[aps,prb,reprint,superscriptaddress]{revtex4-2}

\usepackage{amsmath, amssymb}
\usepackage{bm}
\usepackage{graphicx}
\usepackage{color}
\usepackage{dcolumn}
\usepackage{bm}
\usepackage[hidelinks]{hyperref}
\hypersetup{
  colorlinks   = true, 
  urlcolor     = blue, 
  linkcolor    = blue, 
  citecolor   = red 
}
\usepackage{multirow}

\allowdisplaybreaks[1]

\begin{document}


\title{Vibronic effect on resonant inelastic x-ray scattering in cubic iridium hexahalides}

\author{Naoya Iwahara}
\email[]{naoya.iwahara@gmail.com}
\affiliation{Graduate School of Engineering, Chiba University, 1-33 Yayoi-cho, Inage-ku, Chiba-shi, Chiba 263-8522, Japan} 

\author{Wataru Furukawa}
\affiliation{Department of Materials Science, Faculty of Engineering, Chiba University, 1-33 Yayoi-cho, Inage-ku, Chiba-shi, Chiba 263-8522, Japan}

\date{\today}

\begin{abstract}
In resonant inelastic x-ray scattering (RIXS) spectra of K$_2$IrCl$_6$, the peak for the $j=3/2$ multiplet states shows a splitting that resembles non-cubic crystal-field effect although the compound is cubic down to 0.3 K.  
Here we theoretically describe the RIXS spectra concomitantly treating the spin-orbit and vibronic interactions.
We found that the dynamic Jahn-Teller effect in the $j=3/2$ multiplet states gives rise to the splitting of the RIXS spectra and the broadening of the spectra in raising the temperature.
The validity of the interaction parameters for the simulations is supported by our {\it ab initio} calculations. 
Our results suggest that, in cubic iridium compounds, the dynamic Jahn-Teller effect induces the splitting of RIXS spectra without lowering the symmetry. 
\end{abstract}

\maketitle

\section{Introduction}
Strong spin-orbit coupling gives rise to rich quantum phenomena in $4d/5d$ transition metal compounds \cite{Takagi2019, Takayama2021, Trebst2022}.
One of the most investigated phenomena is the Kitaev spin liquid phase in Mott-insulating honeycomb compounds \cite{Jackeli2009}. 
The candidate honeycomb materials consist of edge-sharing octahedra containing heavy $d^5$ ions. 
The spin-orbit coupling of the $d^5$ ion in the octahedral environment makes the local quantum states spin-orbit entangled $j=1/2$ type and the exchange interaction between the $d^5$ sites bond-dependent (Kitaev-type anisotropy). 
The Kitaev exchange interaction appears in other lattices such as triangular and face centered cubic (fcc) lattices \cite{Jackeli2009}, which could induce various magnetic phases \cite{Kimchi2014, Cook2015}.

Motivated by the theoretical predictions, much experimental effort has been made to search Kitaev and related systems, while realizing pure $j=1/2$ states in materials is far from trivial.
Contrary to the assumption of a perfect octahedral environment in theories, candidate materials exhibit symmetry lowering, resulting in the hybridization of the $j=1/2$ and the excited $j=3/2$ states.

Recently, K$_2$IrCl$_6$ was reported as the first iridium compound with pure $j=1/2$ states on metal sites \cite{Khan2019}. 
The neutron powder diffraction data show that the compound is cubic down to 0.3 K \cite{Reigiplessis2020}.
The $g$-tensor measured by electron paramagnetic resonance is consistent with the $g$ for $j=1/2$ in an octahedral environment, suggesting no symmetry lowering \cite{Bhaskaran2021}.

Nevertheless, spectroscopic data look controversial to the cubic structure of K$_2$IrCl$_6$. 
Within the octahedral environment, the excited states of a single Ir site are four-fold degenerate $j=3/2$ multiplet. 
In contrast, the resonant inelastic x-ray scattering (RIXS) measurement of the $j=3/2$ states shows a splitting of $\approx 50$ meV down to low temperature (10 K) \cite{Reigiplessis2020}. 
The RIXS spectra of cubic K$_2$IrBr$_6$ (in high-temperature phase) and Cs$_2$IrCl$_6$ resemble those of K$_2$IrCl$_6$ too \cite{Reigiplessis2020, Khan2021, Wei2023}. 
The Raman scattering spectra of K$_2$IrCl$_6$ also show a similar splitting in the same range of energy \cite{Lee2022}. 
As the origin of the splitting, non-cubic deformation of each octahedron and crystal-field splitting by rotational modes of the octahedra were discussed, whereas both interpretations require symmetry lowering of the system. 
Moreover, the latter leads to the splitting of only a few meV according to the previous experiments and calculations \cite{Khan2021}.

We recently proved {\it ab initio} that the vibronic excitations induce the splittings of peaks in the RIXS spectra of a similar compound, K$_2$RuCl$_6$ \cite{Takahashi2021, Iwahara2023}.  
In this compound, the dynamic Jahn-Teller (JT) effect develops in the excited spin-orbit multiplet states and determines the shape and temperature evolution of the RIXS spectra. 
The dynamic JT effect is characterized by the superpositions of many electron-lattice configurations with JT-deformed structures, and hence, structural anisotropy does not appear in K$_2$RuCl$_6$. 
Since the vibronic coupling tends to be stronger in $5d$ systems than in $4d$ systems \cite{Iwahara2018}, the dynamic JT effect would give nonnegligible changes in the RIXS spectra of the Ir compounds.

In this work, we examine the presence of the dynamic JT effect and reveal its impact on the RIXS spectra of the Ir$^{4+}$ compounds. 
We calculate the vibronic states of the dynamic JT model and, based on them, the RIXS spectra. 
We confirm the validity of the derived interaction parameters by a series of post Hartree-Fock calculations.

\section{Vibronic model for $5d^5$ ion}
Here we explain the theoretical framework for the quantum states and the RIXS spectra of a single Ir site in the cubic K$_2$Ir$X_6$ ($X =$ Cl, Br).
We discuss only the phenomena in a single Ir octahedron because the energy scale of the intersite exchange interaction (about 1 meV) \cite{Khan2019, Bhaskaran2021, Khan2021} is several tens times smaller than the splitting (about 35-60 meV) in the RIXS spectra \cite{Reigiplessis2020, Khan2021}.

The quantum states of each $5d^5$ site are determined by the interplay of the ligand field, spin-orbit, and vibronic interactions. 
We consider the interactions in the order because their energy scales decrease from several eV by about one order of magnitude. 
The octahedral ligand field splits the $5d$ orbitals into $e_g$ doublet and $t_{2g}$ triplet. 
The $t_{2g}$ orbitals are lower in energy and have five electrons (one hole).
Hereafter, we use the hole picture. 
Within the $t_{2g}$ orbitals, the spin-orbit coupling is unquenched \cite{Sugano1970}: 
\begin{align}
 \hat{H}_\text{SO} &= -\lambda \tilde{\bm{l}} \cdot \hat{\bm{s}}, 
 \label{Eq:HSO}
\end{align}
where $\tilde{\bm{l}}$ is the effective $\tilde{l}=1$ orbital operator, $\hat{\bm{s}}$ is the spin operator, and $\lambda$ $(>0)$ is the spin-orbit coupling parameter. 
The $t_{2g}$ hole couples to the JT active $E_g$ vibrations too:
the JT model including the harmonic oscillator Hamiltonian for the $E_g$ modes is \cite{Bersuker1989}
\begin{align}
 \hat{H}_\text{JT} &= 
 \sum_{\gamma=u,v} \frac{\hslash \omega}{2} \left( \hat{p}_\gamma^2 + \hat{q}_\gamma^2 \right)  
 - \hslash \omega g \left[
   \left( -\frac{1}{2}\hat{q}_u + \frac{\sqrt{3}}{2} \hat{q}_v \right) \hat{P}_{yz} 
   \right.
   \nonumber\\
 &+ 
 \left.
 \left( -\frac{1}{2}\hat{q}_u - \frac{\sqrt{3}}{2} \hat{q}_v \right) \hat{P}_{zx} 
 + \hat{q}_u \hat{P}_{xy} 
 \right].
 \label{Eq:HJT}
\end{align}
Here $\hat{q}_\gamma$ are the dimensionless normal coordinates for the $E_g$ modes ($\gamma = u, v$, where $u$ and $v$ transform as $z^2$ and $x^2-y^2$, respectively), 
$\hat{p}_\gamma$ the conjugate momenta of $\hat{q}_\gamma$, 
$\omega$ the frequency of the $E_g$ modes, 
$g$ the dimensionless vibronic coupling parameter,
and $\hat{P}_\gamma$ ($\gamma = yz, zx, xy$) is the projection operator into the $t_{2g}\gamma$ orbital. 
We ignore the vibronic coupling to the $T_{2g}$ modes because it is much weaker than that to the $E_g$ modes.

The spin-orbit interaction splits the $t_{2g}^1$ hole configurations into effective $j = 1/2$ and $j=3/2$ multiplet states. 
These states belong to the eigen energies of $\hat{H}_\text{SO}$, $-\frac{\lambda}{2}$ and $\lambda$, respectively.
Using the spin-orbit multiplet states, the vibronic coupling part of $\hat{H}_\text{JT}$ becomes 
\begin{align}
 \frac{\hslash \omega g}{2} 
 \begin{pmatrix}
  0 & 0 & 0 & \sqrt{2} \hat{q}_v & 0 & -\sqrt{2} \hat{q}_u \\
  0 & 0 & \sqrt{2} \hat{q}_u & 0 & -\sqrt{2} \hat{q}_v & 0 \\
  0 & \sqrt{2} \hat{q}_u & -\hat{q}_u & 0 & -\hat{q}_v & 0 \\
  \sqrt{2} \hat{q}_v & 0 & 0 & \hat{q}_u & 0 & -\hat{q}_v \\
  0 & -\sqrt{2} \hat{q}_v & -\hat{q}_v & 0 & \hat{q}_u & 0 \\
  -\sqrt{2} \hat{q}_u & 0 & 0 & -\hat{q}_v & 0 & -\hat{q}_u \\
 \end{pmatrix}.
 \label{Eq:VJT}
\end{align}
The basis of the matrix is in the order of $j=\frac{1}{2}$ ($m_j = -\frac{1}{2}, \frac{1}{2}$) and $j = \frac{3}{2}$ ($m_j = -\frac{3}{2}, -\frac{1}{2}, \frac{1}{2}, \frac{3}{2}$).
Eq. (\ref{Eq:VJT}) indicates the presence of the JT effect in the $j = 3/2$ multiplet states and the pseudo JT effect between the $j=1/2$ and $j=3/2$ spin-orbit multiplets.

The energy eigenstates of $5d^5$ sites take the form of spin-orbit and lattice entanglement.
Since the vibronic interaction, Eq. (\ref{Eq:VJT}), hybridizes the spin-orbit multiplet states via the quantized normal coordinates, the energy eigenstates (vibronic states) of $\hat{H}_\text{SO} + \hat{H}_\text{JT}$ are 
\begin{align}
 |\psi_\nu\rangle = \sum_{jm_j} |jm_j\rangle \otimes |\chi_{jm_j; \nu}\rangle, 
 \label{Eq:vibronic}
\end{align}
where $|\chi_{jm_j;\nu}\rangle$ include the information of the lattice degrees of freedom. 
The vibronic states reduce to the direct products of the spin-orbit multiplet and vibrational states in the limit of $g \rightarrow 0$.

Now, we turn to the Ir $L_3$ RIXS spectra of the vibronic system. 
The $L_3$ RIXS is a photon-in photon-out process involving electron transitions from the $2p_{3/2}$ states to the $t_{2g}$ ($5d$) orbitals, and then, from the $t_{2g}$ orbitals to the $2p_{3/2}$ within an Ir ion. 
Assuming that the initial and final states of the system are of vibronic type, Eq. (\ref{Eq:vibronic}), the cross-section for the $L_3$ RIXS process in a single Ir center is \cite{Iwahara2023}
\begin{align}
 \frac{d^2\sigma}{d\Omega dk'} &\propto
 \sum_{\nu}
 \rho_\nu
 \frac{k'}{k}
 \left|
   \sum_{\alpha'\alpha}
   \epsilon_{\bm{k}'\lambda',\alpha'}
   \epsilon_{\bm{k}\lambda,\alpha}
   \langle \nu'|\hat{d}_{\alpha'} \hat{P}_{\text{eh}} \hat{d}_\alpha|\nu\rangle 
 \right|^2
 \nonumber\\
 &\times
 \delta\left(E_{\nu} + \hslash \omega - E_{\nu'} - \hslash \omega'\right),
 \label{Eq:crosssection}
\end{align}
by using the Kramers-Heisenberg formula \cite{Sakurai1967, RIXS} and the dipole and fast collision approximations \cite{Luo1993, vanVeenendaal2006}.
In Eq. (\ref{Eq:crosssection}), $\nu$ ($\nu'$) indicate the initial (final) vibronic states, 
$\rho_\nu$ the canonical distribution for the Ir center, 
$\hat{d}_\alpha$ the $\alpha$ ($=x,y,z$) component of the electric dipole operator, 
$\hat{P}_\text{eh}$ the projection operator into the electron-hole states with $(2p_{3/2})^3(t_{2g})^6$ electron configurations,
and $\hslash \omega_i$ ($\hslash \omega_f$) and $\bm{\epsilon}_{\bm{k}\lambda}$ ($\bm{\epsilon}_{\bm{k}'\lambda'})$ are, respectively, the energy and the polarization of the incident (scattered) photon.
We used the fast collision approximation for the Ir compounds because the energy scale of the lifetime broadening of Ir ion ($\approx$ 4-5 eV) \cite{Clancy2012} is one order of magnitude larger than that of the electron-hole dynamics measured by the $j=3/2$ band width ($\approx 0.7$ eV) \cite{Khan2019}.

\section{Computational methods}
\label{Sec:computation}
For the derivation of the vibronic states, Eq. (\ref{Eq:vibronic}), we performed a numerical diagonalization of the model Hamiltonian [see e.g. Refs. \cite{Iwahara2023, Iwahara2018, Iwahara2017}]. 
We expanded the lattice parts of the vibronic states, Eq. (\ref{Eq:vibronic}), as 
\begin{align}
 |\chi_{jm_j; \nu}\rangle &= \sum_{n_u, n_v} |n_u, n_v\rangle C_{jm_j, n_u, n_v; \nu}.
\end{align}
To determine the coefficients $C_{jm_j, n_u, n_v; \nu}$, we calculated the Hamiltonian matrix of the JT model ($\hat{H}_\text{SO} + \hat{H}_\text{JT}$) and diagonalized it. 
The basis of the matrix was a set of the direct products of the $|j, m_j\rangle$ ($j = \frac{1}{2}, \frac{3}{2}$) and the eigenstates of the harmonic oscillator part of $\hat{H}_\text{JT}$, $|n_u, n_v\rangle$ $(n_u, n_v = 0, 1, 2, \cdots)$. 
We truncated our basis so that $0 \le n_u + n_v \le 15$.

The dynamic JT Hamiltonian contains interaction parameters, $\lambda$, $\omega$, and $g$.
We took $\lambda$ from the RIXS data and $\omega$ from the Raman scattering data and treated $g$ as a variable. 
We determined $g$ so that we can reproduce the experimental RIXS spectra.

To check the validity of the chosen parameters, we also derived all the interaction parameters by a series of post Hartree-Fock calculations as in Refs. \cite{Iwahara2017, Iwahara2018, Iwahara2023}. 
We made clusters of K$_2$IrCl$_6$ and K$_2$IrBr$_6$ using the neutron diffraction data at 0.3 K for the former and 200 K for the latter, respectively \cite{Reigiplessis2020}.
Each cluster consists of three groups of atoms. 
The first group corresponds to K$_8$Ir$X_6$ ($X = $ Cl, Br) cluster, which we treated {\it ab initio} with atomic-natural-orbital relativistic-correlation consistent-valence quadruple-$\zeta$ polarization basis set. 
The second group contains the nearest 12 octahedra and the neighboring 48 K atoms, which are replaced by {\it ab initio} embedded potential \cite{Seijo1999}. 
The third part consists of point charges.
The total charge of the entire cluster is neutral. 
We calculated ${}^2T_{2g}$ states of the clusters with the complete active space self-consistent field (CASSCF) method \cite{Roos2016} followed by a correction within the extended multistate complete active space second-order perturbation theory (XMS-CASPT2) \cite{Granovsky2011, Shiozaki2011}, and then, 
the spin-orbit $j$ multiplet states with spin-orbit restricted active space state interaction (SO-RASSI) method \cite{Roos2016} implemented in {\tt OpenMolcas} \cite{molcas1_short, molcas2_short}. 
In these calculations, we set three $5d$ $t_{2g}$ orbitals with one hole as the active space.
We determined $\lambda$ by using the SO-RASSI energy gap.

For the calculations of the vibronic coupling parameters and the frequencies, we fit the ${}^2T_{2g}$ adiabatic potential energy surfaces with respect to the JT deformation to the JT model, Eq. (\ref{Eq:HJT}) [See Refs. \cite{Iwahara2017, Iwahara2018, Iwahara2023}]. 
We generated several JT-deformed Cartesian coordinates of an octahedron by using $\bm{R}_A = \bm{R}_A^{(0)} + Q_u \left(\bm{e}_{Eu}\right)_A/\sqrt{M}$ and calculated the term energies $U_\gamma$ ($\gamma = yz, zx, xy$) at each structure. 
Here $\bm{R}_A$ are the Cartesian coordinates of ligand atom $A$, $\bm{R}^{(0)}$ indicates the experimental octahedral configurations, $\bm{e}_{Eu}$ is the JT-active $Eu$ polarization vectors of the neighboring six Cl's or Br's, $Q_u$ the mass-weighted normal coordinates ($Q_u = \sqrt{\hslash/\omega} q_u$), and $M$ the mass of ligand atoms. 
The gradient of $U_{xy}$ with respect to $Q_u$ is the vibronic coupling parameter, $V = -\partial U_{xy}/\partial Q_u$. 
The curvature of $U$ is mainly $\omega^2$ with a small correction from the quadratic vibronic coupling.
The quadratic vibronic coupling has a similar form as the linear one in Eq. (\ref{Eq:HJT}): the former is obtained by replacing $\hat{q}_\gamma$ and $g$ in the latter with the symmetrized quadratic form of $\hat{q}$ [$\hat{q}_u \rightarrow -\frac{1}{\sqrt{2}}(\hat{q}_u^2-\hat{q}_v^2)$ and $\hat{q}_v \rightarrow \sqrt{2}\hat{q}_u\hat{q}_v$] and $w$, respectively (see e.g., Ref. \cite{Iwahara2018}). 
With $V$ and $\omega$, $g$ is defined by $g = V/\sqrt{\hslash \omega^3}$.


\begin{figure*}[tb]
 \begin{tabular}{lll}
  (a) &~& (b) \\
  \includegraphics[width=0.47\linewidth]{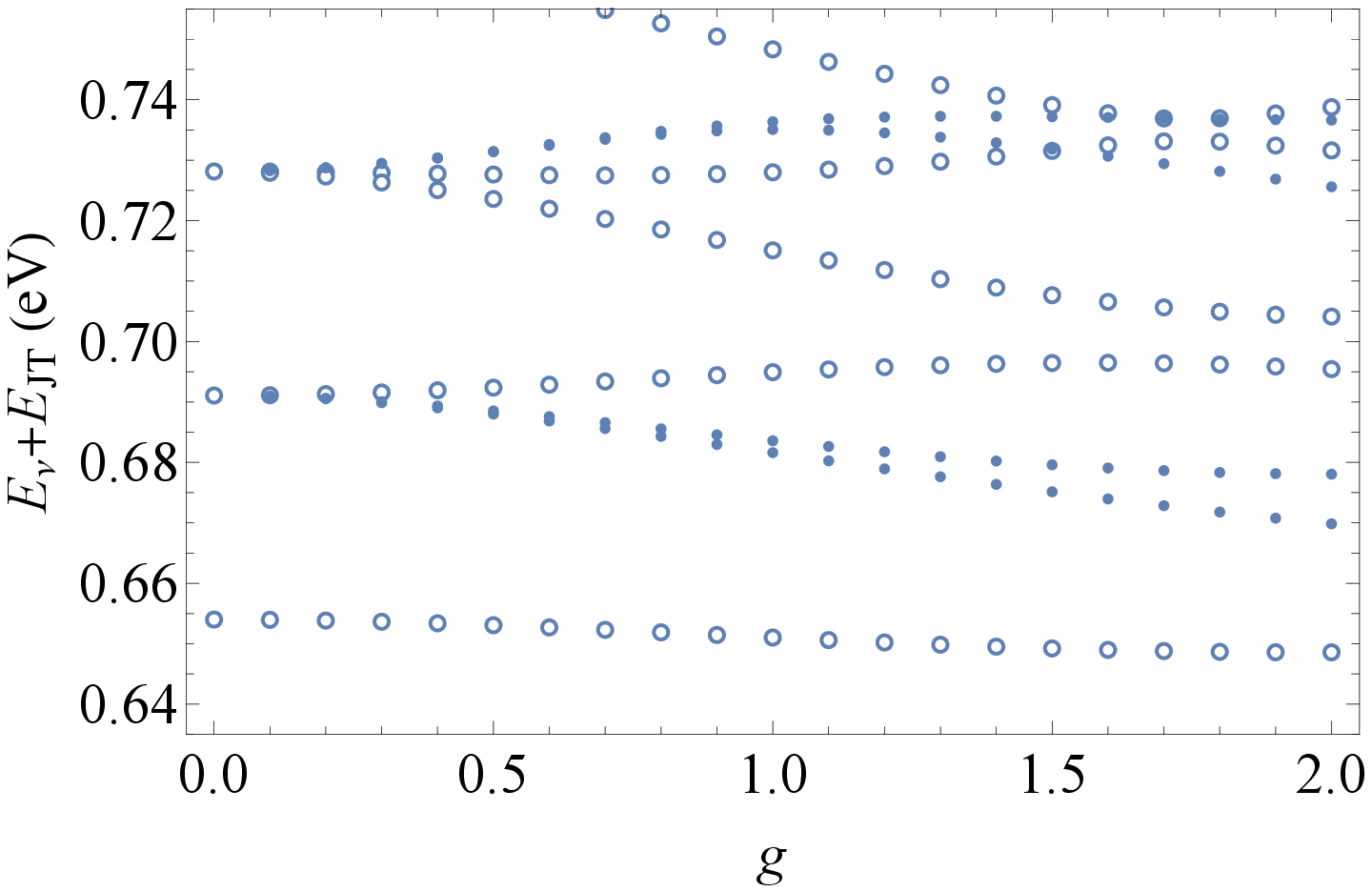}
  &&
  \includegraphics[width=0.47\linewidth]{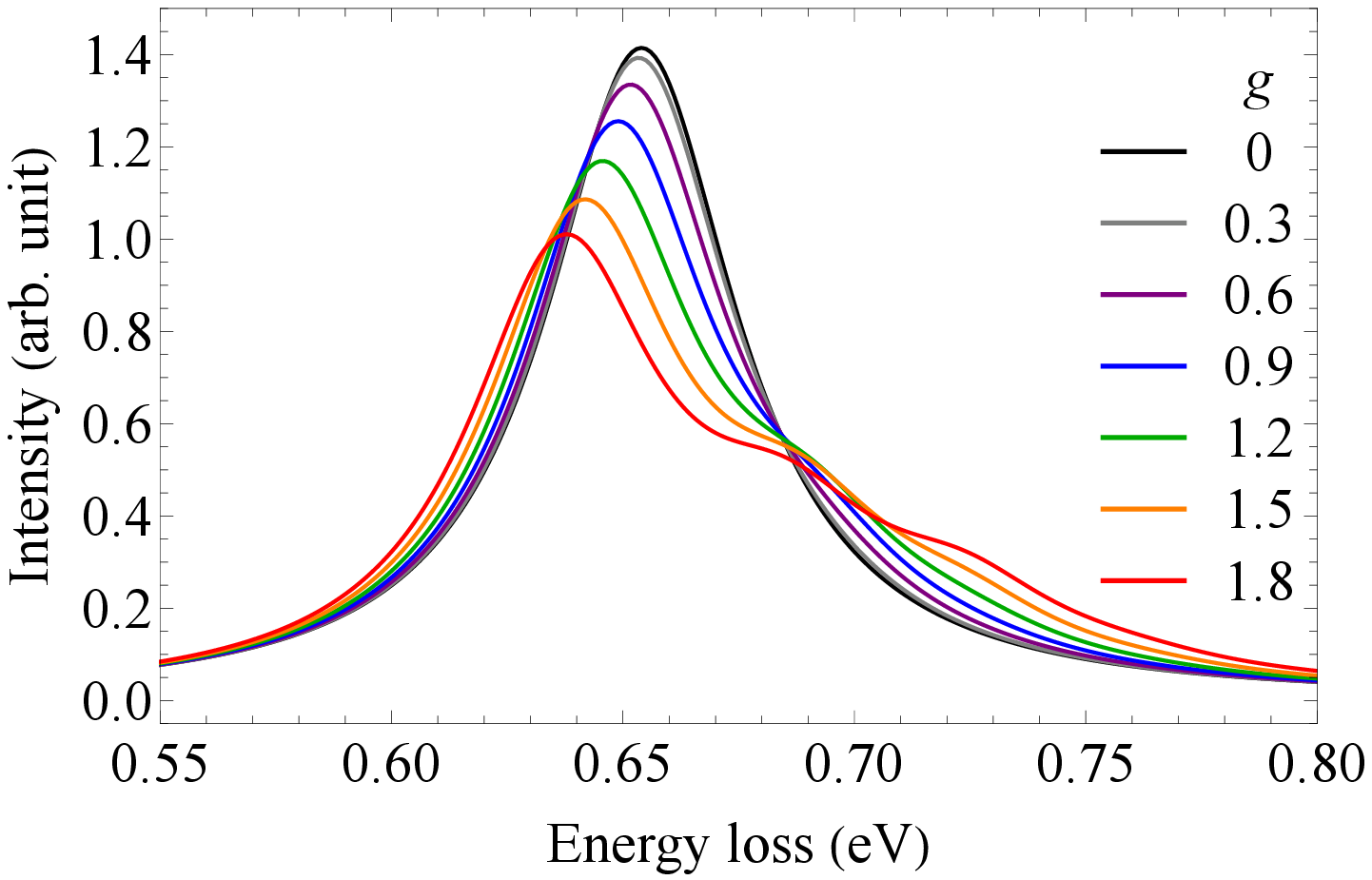}
  \\
  (c) && (d) \\
  \includegraphics[width=0.47\linewidth]{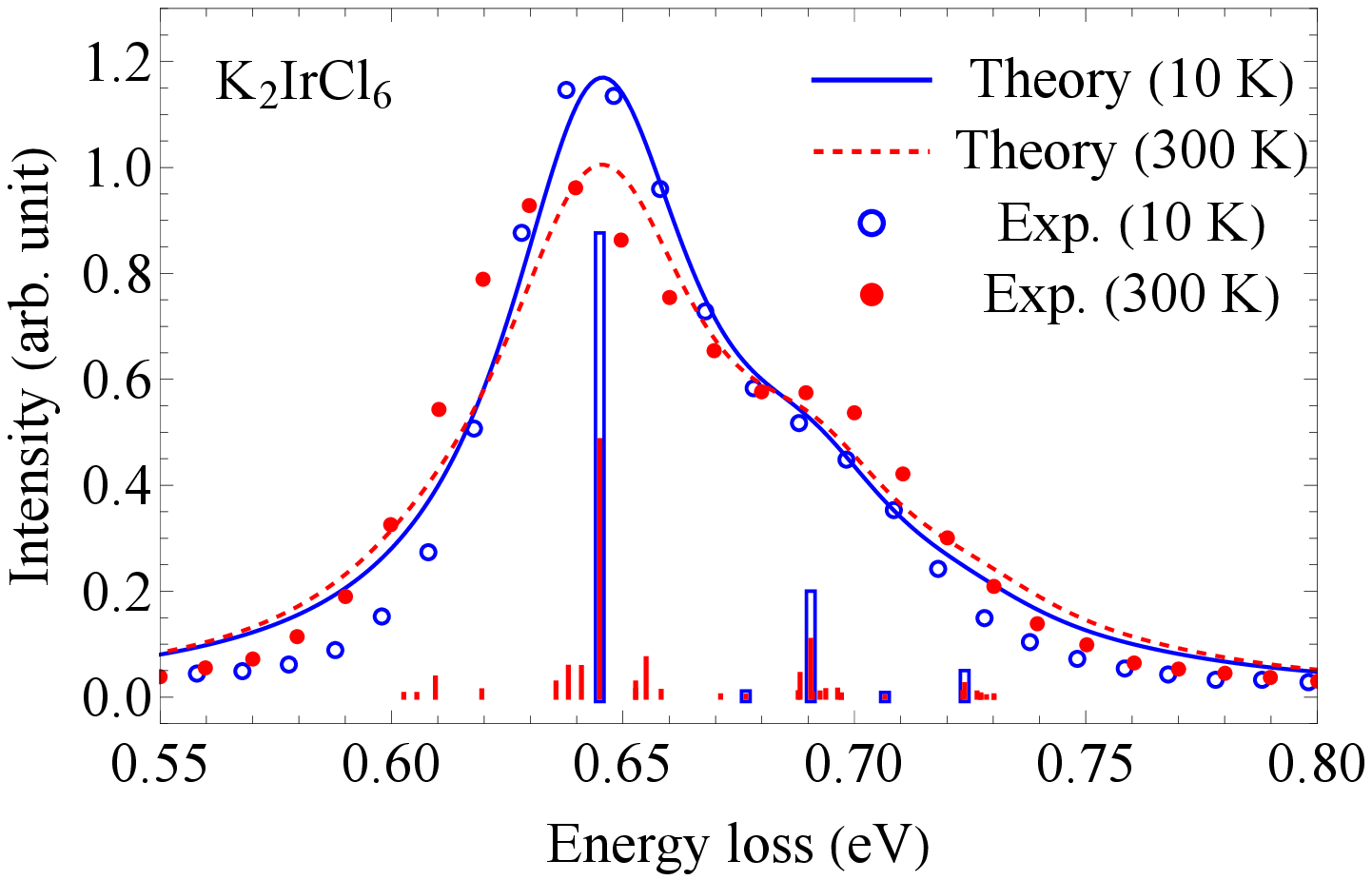}
  &&
  \includegraphics[width=0.47\linewidth]{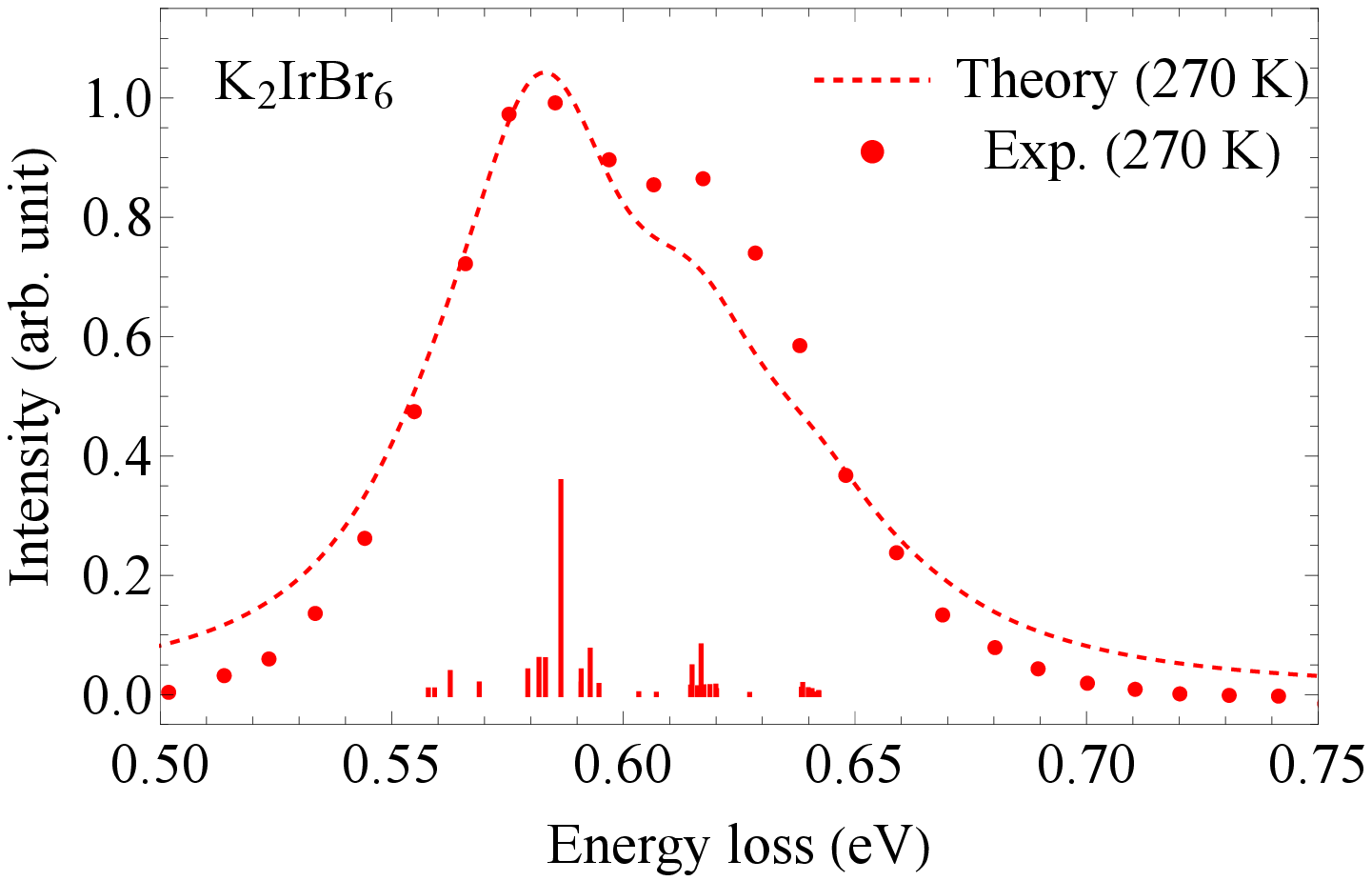}
 \end{tabular}
 \caption{
   Vibronic levels and RIXS spectra for (a)-(c) K$_2$IrCl$_6$ and (d) K$_2$IrBr$_6$. 
   (a) The vibronic energy spectra with respect to $g$.  $E_\text{JT} = \hslash \omega (g/2)^2/2$.
   (b) The RIXS spectra with respect to $g$ at $T = 10$ K. 
   (c) Theoretical ($g=1.2$) and the experimental \cite{Reigiplessis2020} RIXS spectra at $T = 10$ K and $T = 300$ K. 
   The vertical lines indicate the cross-sections ($\times 10$) at 10 K (blue) and at 300 K (red).
   (d) Theoretical ($g = 1.5$) and the experimental \cite{Khan2021} RIXS spectra at $T = 270$ K.
   }
   \label{Fig:calc}
\end{figure*}

\section{Results and discussions}
\subsection{Vibronic energy spectra}
\label{Sec:vibronic}
We calculated the vibronic levels of the dynamic JT model for a single Ir$^{4+}$ site.
For the calculations, we used the interaction parameters from the experiments of K$_2$IrCl$_6$: $\lambda = 0.436$ eV \cite{Reigiplessis2020} and $\omega = 37.1$ meV \cite{Lee2022}.
We reduced the reported $\lambda$ by 0.04 eV for a better agreement between the theoretical and experimental RIXS spectra (Sec. \ref{Sec:RIXS}).

The vibronic energy spectra around the excited spin-orbit $j=3/2$ multiplet level show nontrivial distribution. 
Figure \ref{Fig:calc}(a) indicates the energy spectra with respect to $g$. 
Each level is either two-fold (the points) or four-fold (the open circles) degenerate. 
At $g = 0$ (no JT effect), the energy eigenstates in the figure correspond to the product states of $|j=3/2, m_j\rangle \otimes |n_u, n_v\rangle$ type ($n_u+n_v = 0, 1, 2, ...$). 
Turning on $g$, the vibronic coupling hybridizes these product states as Eq. (\ref{Eq:vibronic}). 
Consequently, the distribution of the vibronic levels for $|g|>0$ differs from that without the JT effect ($g = 0$).
Although the vibronic coupling hybridizes the ground $j=1/2$ states with the excited $j=3/2$ states, the ground states remain close to the $j=1/2$ multiplet states, $|j=1/2, m_j\rangle \otimes |0,0\rangle$, in the present range of $g$: the weight (probability) of the pure $j=1/2$ components in the ground state is 0.998 at $g = 1.0$ and 0.990 at $g = 2$.

We also calculated the vibronic states of K$_2$IrBr$_6$. 
We used $\lambda = 0.395$ eV \cite{Khan2021} and $\omega = 24.7$ meV.
We reduced $\lambda$ from the experimental value by 0.05 eV.
We estimated $\omega$ by assuming that the elastic couplings $K$ for the $E_g$ modes are common for K$_2$IrCl$_6$ and K$_2$IrBr$_6$ and using $\omega \propto \sqrt{K/M}$.
The pattern of the calculated vibronic states of K$_2$IrBr$_6$ resembles that of K$_2$IrCl$_6$. 
The ground state is well described with $|j=1/2, m_j\rangle \otimes |0,0\rangle$: the weight 0.999 at $g = 1.0$ and 0.995 at $g = 2.0$.

\subsection{Vibronic peaks in RIXS spectra} 
\label{Sec:RIXS}
With the derived vibronic states, we calculated the RIXS spectra around the $j=3/2$ multiplet level at 10 K using Eq. (\ref{Eq:crosssection}).
We set the incident and scattered photons following the experimental condition \cite{Reigiplessis2020}.
We convoluted the cross-section, Eq. (\ref{Eq:crosssection}), with the Lorentzian function with width $\Gamma = 50$ meV.

Varying $g$, the shape of the RIXS spectrum changes [Fig. \ref{Fig:calc}(b)].
When $g = 0$, the RIXS spectrum has only one peak corresponding to the gap between the ground $j=1/2$ level and the $j=3/2$ level with no vibrational excitations ($n_u=n_v=0$).
The transitions from the ground $j=1/2$ states [$|j=1/2, m_j\rangle \otimes |0,0\rangle$] to the vibrationally excited $j=3/2$ multiplets [$|j=3/2, m_j\rangle \otimes |n_u, n_v\rangle$, $n_u+n_v>0$] are zero due to the orthogonality of the vibrational states. 
As $g$ increases, the peak splits into a few originating from the transitions between the ground $j=1/2$ states 
and several excited vibronic states (see the blue vertical lines): 
the latter states contain non-negligible contributions from $|j=3/2, m_j\rangle \otimes |0,0\rangle$ states.

The calculated RIXS spectrum with $g = 1.2$ agrees well with the experimental one from Ref. \cite{Reigiplessis2020} [Fig. \ref{Fig:calc}(c)].
The calculated spectrum has two strong peaks at about 0.645 eV and 0.690 eV corresponding to the transitions between the ground $j=1/2$ multiplet and the four-fold degenerate vibronic states [The lowest two four-fold degenerate levels in Fig. \ref{Fig:calc}(a)]. 
The gap between the peaks is about 45 meV, which is close to the experimental estimation of 48 meV.

Raising the temperature from 10 K to 300 K, the shape of the RIXS spectrum changes. 
To calculate the RIXS spectrum at 300 K, we only vary the temperature in the convoluted cross-section: the other parameters remain the same. 
The height of the main peak at 0.645 eV decreases, and the width increases as the temperature rises. 
These temperature dependencies agree with those of the experimental data.
The temperature evolution of the spectra occurs due to the transitions from the thermally populated states with vibrational excitations to the vibronic states [the red lines in Fig. \ref{Fig:calc}(c)]. 
Increasing the temperature from 10 K to 300 K, the statistical weight of the ground $j=1/2$ states reduces from 1.000 to 0.558, which lowers the height of the main peak. 
The slight enhancement of the second peak originates from the transitions between the thermally populated first excited states to the excited vibronic states.

We also calculated the RIXS spectrum of cubic K$_2$IrBr$_6$ at 270 K.
For the calculation, we used the vibronic states with $g = 1.5$ and the Lorentzian function with $\Gamma =$ 40 meV.
We estimated $g$ for K$_2$IrBr$_6$ from $g$ for K$_2$IrCl$_6$ by using $g \propto M^{1/4}$. 
The relation is obtained by using the definition of $g = V/\sqrt{\hslash\omega^3}$ and $V, \omega \propto M^{-1/2}$.
Figure \ref{Fig:calc}(d) shows a good agreement between the calculated and experimental RIXS spectra \cite{Khan2021}.
The calculated spectrum has two peaks at 0.587 eV and 0.617 eV, and the gap between them (30 meV) is smaller than that of K$_2$IrCl$_6$, which is in line with the experimental observation (the gap is 33 meV) \cite{Khan2021}.
The theoretical spectrum underestimates the height of the peak at 0.617 eV. 
The heights of the peaks could vary by a hybridization of the corresponding $j=3/2$ vibronic states due to the intersite interactions in the intermediate states. 
The gap between the peaks and the intersite interactions are smaller \cite{Reigiplessis2020, Khan2021} and larger \cite{Khan2019} in K$_2$IrBr$_6$ than in K$_2$IrCl$_6$, respectively, and thus, the hybridization effect should be enhanced in K$_2$IrBr$_6$.

Our calculations suggest that the ground states are close to the pure spin-orbit $j=1/2$ multiplet states in K$_2$IrCl$_6$ down to low-temperature and K$_2$IrBr$_6$ at high-temperature. 
The weight of the $j=1/2$ multiplets in each of the ground vibronic doublets is 0.997 in K$_2$IrCl$_6$ and 0.998 in K$_2$IrBr$_6$.

\begin{figure}[tb]
\includegraphics[width=0.9\linewidth]{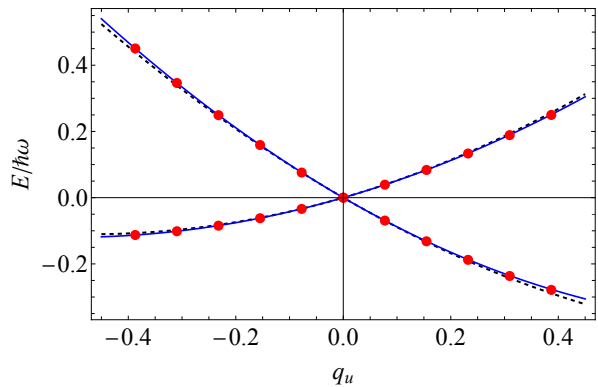}
\caption{The $^2T_{2g}$ term energies of K$_2$IrCl$_6$ with respect to the $E_gu$ deformations.
The red points are the {\it ab initio} data, and the solid (blue) and dashed (black) lines are the model adiabatic potential energy surfaces with and without quadratic vibronic coupling, respectively. 
}
\label{Fig:abinitio}
\end{figure}

To confirm the validity of our interaction parameters, we performed {\it ab initio} calculations [Fig. \ref{Fig:abinitio}].
The calculated interaction parameters are $\lambda = 0.469$ eV, $\omega =$ 40.7 meV, $g = 0.94$, and $w = 0.11$ for K$_2$IrCl$_6$ and $\lambda = 0.472$ eV, $\omega = 24.5$ meV, $g = 1.03$, and $w = 0.13$ for K$_2$IrBr$_6$.
The calculated $\lambda$ and $\omega$ are close to the experimental values: the calculated data are about 10 \% larger than the experimental data \cite{Reigiplessis2020, Khan2021, Lee2022}. 
The $g$ for K$_2$IrBr$_6$ is larger than that of K$_2$IrCl$_6$, which is in line with our estimation above. 
The overestimation of $\lambda$ and $\omega$ is due to insufficient delocalization of the $t_{2g}$ orbitals over the ligands within the post Hartree-Fock method. 
Due to the overestimation of $\omega$'s, the $g$'s are underestimated. 
The calculated $w$ is about 10 times weaker than $g$, which allows neglecting the quadratic vibronic coupling in the JT model, Eq. (\ref{Eq:HJT}).
Overall, we conclude that the {\it ab initio} calculations support the validity of the parameters extracted from the experimental data and the mechanism of the splitting.

Based on our interaction parameters, we rule out the non-cubic (or static JT effect) scenario. 
The JT deformations that reproduce the splitting of the $j=3/2$ peak correspond to the shifts of the ligand atoms of about 0.007 {\AA} for K$_2$IrCl$_6$ and 0.004 {\AA} for K$_2$IrBr$_6$. 
Such large shifts were not observed in the x-ray and neutron diffraction data of K$_2$IrCl$_6$ down to low temperature.

We propose that the rotational modes of the octahedra enlarge the line width of the RIXS spectra. 
The rotational modes quadratically couple to the $t_{2g}$ orbitals \cite{Khan2021} and, hence, they can contribute to the spin-orbit and lattice entanglement as the JT active modes.
The rotational modes will give rise to new transitions between the levels around $j=1/2$ and $j=3/2$ with about one order of magnitude smaller gaps than the splitting of the $j=3/2$ peak according to their frequencies: within the present resolution of energy, these transitions will enlarge the line width of the spectra.  
Due to the enlargement of the line width by raising the temperature, the peak positions of the spectra will shift, and hence, the energy gap will slightly increase.

Finally, we comment on the nature of the JT effect in the $4d$ and $5d$ spin-orbit Mott insulators.
Our analysis shows that the vibronic coupling in K$_2$IrCl$_6$ ($5d^5$) is only about 10 \% larger than that in K$_2$RuCl$_6$ ($4d^4$) \cite{Iwahara2023}: the change is much smaller than that of spin-orbit coupling.
Under this situation, the $4d$ and $5d$ compounds should exhibit different regimes of the interplay of the JT effect and spin-orbit coupling. 
Emergent phenomena induced by the interplay of the vibronic and spin-orbit couplings deserve future studies.

\section{Conclusion}
In this work, we theoretically demonstrated that the dynamic JT effect determines the shape and the temperature dependence of the RIXS spectra of K$_2$IrCl$_6$. 
In particular, the dynamic JT effect in the $j=3/2$ multiplet states develops and splits the peak for the $j=3/2$ multiplet in the RIXS spectra. 
The interaction parameters used for the simulations are consistent with our {\it ab initio} interaction parameters. 
Thus, our results indicate that the system is cubic. 
The present results suggest that the RIXS spectra of spin-orbit Mott insulators with (quasi) degenerate multiple states on sites have to be interpreted fully considering the dynamic (pseudo) JT effect.

\begin{acknowledgments}
This work was partly supported by the Iketani Science and Technology Foundation and Grant-in-Aid for Scientific Research (Grant No. 22K03507) from the Japan Society for the Promotion of Science.
\end{acknowledgments}

\bibliography{ref}
\end{document}